\documentstyle[preprint,eqsecnum,aps]{revtex}
\begin{document}
 \draft
\preprint{SNUTP 96-084
}
\bigskip
\title{ Color-octet mechanism in \\ $J/\psi$ productions  }
\author{Pyungwon Ko 
 \footnote{To be published in 
 the Proceedings of the KPS-JKPS Winter School held at Seoul, Korea during 
 Feb. 21-28, 1996.} 
\thanks{ pko@phyb.snu.ac.kr} \\ }
\address{Dep. of Physics, Hong-Ik University \\ Seoul 121-791, Korea \\
}
\date{\today}
\maketitle
\begin{abstract}
I review recent developments in theoretical understandings of $J/\psi$
productions in various processes such as $p \bar{p} \rightarrow
J/\psi + X$, $B \rightarrow J/\psi +X$  and 
$J/\psi$ photoproductions including the color-octet mechanism. 
\end{abstract}

\tighten


\narrowtext

\section{Introduction} 
\label{sec:intro} 

In the conventional approach, the inclusive $J/\psi$ 
productions in various high energy processes   
have  been studied in the framework of perturbative QCD (PQCD) and the 
color singlet model \cite{pqcd}.  
In this model, one assumes that the color-singlet $c\bar{c}$ state is 
produced in the ${^{3}S_1}$ state in the short distance scale ({\it i.e.} 
via some hard process which is calculable in PQCD) from the
beginning, and this object eventually evolves into the physical $J/\psi$
state in the long distance scale.  However, this approach, 
when applied to the 
$J/\psi$ or $\psi^{'}$ production at the Tevatron, severely underestimates
the productions rate \cite{cdf}.  
In order to reconcile the data and PQCD predictions,
a new mechanism for heavy quarkonium productions has been suggested 
\cite{fleming},  the color-octet gluon fragmentation into $J/\psi$.  
Also, the color-octet mechanism in heavy quarkonium  productions at hadron 
colliders through the color-octet $(c \bar{c})_8$ pair in various partial 
wave  states ${^{2S+1}L_J}$  
has been considered beyond the color-octet gluon fragmentation approach 
\cite{cho}, \cite{cho2}. 
In Refs.~\cite{cho} and  \cite{cho2}, 
a large class of color-octet diagrams has been 
considered which can contribute to the $J/\psi$ production at hadron 
colliders.  Here, the basic picture is the foolowing :
at the parton level, one can have color-octet $c\bar{c}$ states with
various $^{2S+1}L_J$, 
\begin{eqnarray}
q \bar{q} & \rightarrow & (c\bar{c})(^{3}S^{(8)}_{1}), 
\\
g g & \rightarrow & (c\bar{c})({^{1}S^{(8)}_{0}}~{\rm or~} {^{3}P^{(8)}_{J}}),
\end{eqnarray}
at the short distance scale, and the subsequent evolution of the 
$(c\bar{c})_{8}({^{2S+1}L_{J}})$ object into a physical $J/\psi$ by 
absorbing/emitting soft gluons at the long distance scale  \footnote{
This kind of processes is clearly possible, but has never been addressed 
before in a quantitative way. The color evaporation model \cite{evap} 
is in principle close to this picture,
but there are some parameters which are introduced
phenomenologically, and thus have no definitions in QCD. }.
The short distance process can be calculated using PQCD in powers of 
$\alpha_s$, whereas  the long distance part is treated as a new parameter
$\langle 0 | O_{8}^{\psi} ({^{2S+1}L_{J}}) | 0 \rangle$ 
which characterizes the probability that the color-octet $(c\bar{c}) 
({^{2S+1}L_{J}})$ state evolves into a physical $J/\psi$ by 
emitting/absorbing soft gluons \footnote{See Ref.~\cite{bodwin} for 
more details on NRQCD, and definitions of ${\cal O}_{8}^{\psi}$. }. 
By fitting the $J/\psi$ production at the Tevatron using the usual 
color-singlet production and the cascades from $\chi_{c}(1P)$ and the
color-octet contribution,  the authors of Ref.~\cite{cho2} determined 
\begin{eqnarray}
\langle 0 | O_{8}^{\psi} (^{3}S_{1}) | 0 \rangle   &=&   (6.6 \pm 2.1)
\times 10^{-3}~{\rm GeV}^3,
\\
\frac{\langle 0|{\cal O}_{8}^{\psi}({^3P_{0}})|0\rangle}{M_c^2}
     +\frac{\langle 0|{\cal O}_{8}^{\psi}({^1S_{0}})|0\rangle}{3}
&=&(2.2\pm 0.5)\times 10^{-2}~{\rm GeV}^3
\end{eqnarray}
for $M_{c} = 1.48$ GeV.
Although the numerical values of two matrix elements $\langle 0|
{\cal O}_{8}^{\psi}(^3P_0)|0\rangle$ and $\langle 0|{\cal O}_{8}^{\psi}
(^1S_0)|0\rangle$ are not separately known in Eq.~(1.4), one can still
extract some useful information from it. Since both of the color octet 
matrix elements in Eq.~(1.4) are positive definite, one has
\begin{eqnarray}
0 < \langle 0|{\cal O}_{8}^{\psi}({^1S_0})|0\rangle < (6.6 \pm 1.5) 
\times 10^{-2}~{\rm GeV}^3,
\\
0 < { \langle 0|{\cal O}_{8}^{\psi}({^3P_0})|0\rangle \over M_{c}^2} 
<  (2.2 \pm 0.5) \times 10^{-2}~{\rm GeV}^3.
\end{eqnarray}
These inequalities can provide us with some predictions on various 
quantities related with inclusive $J/\psi$ productions in other 
high energy processes, which enables us to test the idea of color-octet
mechanism.

Since the color-octet mechanism in heavy quarkonium production is  a 
new idea proposed in order to resolve the $\psi^{'}$ anomaly at the 
Tevatron, it is 
important to test this idea in other high energy processes with 
inclusive heavy quarkonium productions. Up to now, the following 
processes have been considered : $J/\psi$ production at the Tevatron
and fixed target experiments \cite{cho} \cite{cho2} \cite{fleming1}, 
spin alignment of the color-octet produced 
$J/\psi$ \cite{wise}, the polar angle distribution of the $J/\psi$ in the 
$e^+ e^-$ annihilations into $J/\psi +X$ \cite{chen},
inclusive $J/\psi$ production in 
$B$ meson decays \cite{ko},  the $Z^0$ decays at LEP \cite{cheung} 
\cite{baek}, and 
$J/\psi$ photoproductions at the fixed target experiments as well as 
at HERA \cite{kramer2} \cite{fleming2} \cite{ko2}.  
These processes  also  depend on the aforementioned 
three color-octet matrix elements  in different 
combinations from (1.4).   Thus, one can check if the color-octet mechanism
provides reasonable agreements between PQCD and the experimental data
on inclusive $J/\psi$ production rates from these processes.

In this talk, I review two applications of the idea of the
color-octet $J/\psi$ productions, which I was  working on, 
among many recent applications mentioned above  
\footnote{
The case of $J/\psi$ productions at the Tevatron is covered in detail
by Cho and Leibovich in Refs.~[4,5].}.         
First in Sec.~\ref{sec:bdecay}, I discuss
the $B \rightarrow J/\psi + X$ using the
factorization formula derived in Ref.~\cite{ko}, and find that the
relations (1.5) and (1.6) overestimate the branching ratio for $B \rightarrow
J/\psi + X$, especially for (1.6).   
Then, we discuss the $J/\psi$ photoproductions  in the 
in Sec.~\ref{sec:photo}.  For this process, the singlet contribution from 
$\gamma+g \rightarrow J/\psi+g$ (the $\gamma g$ fusion) has long been known.
And we consider the color-octet subprocesses
\begin{equation}
\gamma + g \rightarrow (c \bar{c})({^{1}S^{(8)}_{0}} ~{\rm or}~ 
{^{3}P^{(8)}_{J=0,2}}),
\end{equation}
which have  not been included in previous studies.   
These color-octet $2 \rightarrow 1$ subprocesses can also contribute to
the $2 \rightarrow 2$ subprocesses through 
\begin{eqnarray}
\gamma+g&\rightarrow&(c\bar{c})(^{1}S^{(8)}_{0}~{\rm or}~^{3}P^{(8)}_{J}) + g,
\\
\gamma+q&\rightarrow&(c\bar{c})(^{1}S^{(8)}_{0}~{\rm or}~^{3}P^{(8)}_{J}) + q.
\\
\gamma + g &\rightarrow & (c \bar{c})({^{2S+1}L^{(8)}_{J}}) + g.
\end{eqnarray}
All of these color-octet $2 \rightarrow 2$ subprocesses are 
calculated in Refs.~\cite{kramer2} \cite{ko2}.
Numerical analyses relevant to the fixed target experiments and HERA 
have been  performed. 
We show that the relations (1.5) and (1.6)
yields too large a cross section for the $J/\psi$ 
photoproduction in the forward direction. They also leads to too rapidly  
growing $d \sigma / dz$ distribution for high $z$ region compared 
to the experimental observations.  
All of these seem to indicate that the relations (1.3) and 
(1.4), especially the latter, are probably overestimated by an order of 
magnitude.  This is not surprising at all, since the analyses in 
Ref. ~\cite{cho2} employed the leading order calculations for the color-singlet
parton subprocess for the $J/\psi$ hadroproduction.
We summarize our review and speculate the origins of these overestimates of 
$J/\psi$ photoproductions and $B$ meson decays in Sec.~\ref{sec:con}.  

\section{Inclusive $J/\psi$ production in $B$ decays}
\label{sec:bdecay}

The effective Hamiltonian for $b \rightarrow c \bar{c} q$ (
with $q = d,s$) is written as \cite{bodwinB}
\begin{eqnarray}
H_{eff} & = & {G_{F} \over \sqrt{2}}~V_{cb} V_{cq}^{*}~
\left[ {{2 C_{+} - C_{-}}
\over 3}~\bar{c} \gamma_{\mu} (1 -\gamma_{5})c~\bar{q} \gamma^{\mu} (1 -
\gamma_{5}) b ~ \right.
\nonumber    \\
 & + & \left. (C_{+} + C_{-})~\bar{c} \gamma_{\mu} (1 - \gamma_{5})
T^{a} c ~\bar{q}\gamma^{\mu} (1 - \gamma_{5}) T^{a} b \right],
\label{eq:heff}
\end{eqnarray}
where $C_{\pm}$'s are the Wilson coefficients at the scale $\mu \approx
M_b$. We have neglected penguin operators, since their Wilson coefficients
are small  and thus they are irrelevant to our case.
To leading order in $\alpha_{s}(M_{b})$ and to all orders in $\alpha_{s}
(M_{b})~{\rm ln}(M_{W}/M_{b})$,  the above Wilson coefficients are
\begin{equation}
C_{+} (M_{b}) \approx 0.87,~~~~~ C_{-} (M_{b}) \approx 1.34.
\end{equation}

According to the factorization theorem for the $S-$wave charmonium
productions in  $B$ decays,  one has
\cite{bodwinB}
\begin{equation}
\Gamma ( b \rightarrow J/\psi + X )  =   { \langle 0 | O_{1}^{J/\psi}
(^{3}S_{1}) | 0 \rangle \over 3 M_{c}^2}~ \hat{\Gamma}_{1}
( b \rightarrow (c \bar{c})_{1} (^{3}S_{1}) + X ),
\end{equation}
in the nonrelativistic limit,
where $\hat{\Gamma}_{1}$ are rates for hard subprocesses of $b$ quark
decaying into a $c \bar{c}$ pair with suitable angular momentum and
vanishing  relative momentum in the color-singlet :
\begin{eqnarray}
\hat{\Gamma}_{1} (b \rightarrow (c\bar{c})_{1}
(^{3}S_{1}) + s,d) & = & (2C_{+} - C_{-}
)^{2} \left( 1 + {8 M_{c}^{2} \over M_{b}^2} \right)~\hat{\Gamma}_{0},
\\
\hat{\Gamma}_{1} (b \rightarrow (c\bar{c})_{1}
(^{1}S_{0}) + s,d) & = &
(2C_{+} - C_{-} )^{2} ~\hat{\Gamma}_{0},
\end{eqnarray}
with
\begin{equation}
\hat{\Gamma}_{0} \equiv |V_{cb}|^{2} \left( {G_{F}^{2} \over 144 \pi}
\right) M_{b}^{3} M_{c} \left( 1 - {4 M_{c}^{2} \over M_{b}^2} \right)^{2}.
\end{equation}
The operator $O_{1}^{H} (^{2S+1}S_{J})$ is defined in terms of heavy
quark field operators in NRQCD \footnote{We follow the notations in
Ref.~\cite{bodwin}, and will not give explicit forms for these dimension-six
operators in this paper.}.
Its  matrix element $\langle 0 | O_{1}^{H} (^{2S+1}S_{J}) | 0 \rangle$
contains the nonperturbative
effects in the heavy quarkonium production processes,
and is proportional to the probability that a $c \bar{c}$ in a color-singlet
$S-$wave state fragments into a  color-singlet $S-$wave $c \bar{c}$ bound
state such as  a  physical  $J/\psi$, or $\psi^{'}$.
It is also  related to the matrix element
$\langle H | O_{1} (^{2S+1}S_{J}) | H \rangle$ and 
the nonrelativistic quarkonium wavefunction as follows :
\begin{equation}
\langle 0 | O_{1}^{J/\psi} (^{3}S_{1}) | 0 \rangle \approx
3~\langle J/\psi | O_{1} (^{3}S_{1}) | J/\psi \rangle
\approx \left(
{9 \over 2 \pi} \right)~  \left| R_{\psi} (0)
\right|^2,
\end{equation}
in the nonrelativistic limit.
Note that dependence on the radial quantum numbers
$n$ enters through the nonperturbative parameters, $\langle 0 | O_{1}^{H}
(^{3}S_{1}) | 0 \rangle$.

Using the leptonic decay width of $J/\psi$ and $\psi^{'}$, one can determine
\begin{eqnarray}
\langle J/\psi | O_{1} (^{3}S_{1}) | J/\psi \rangle & \approx &
2.4 \times 10^{-1}~~{\rm GeV}^{3},
\\
\langle  \psi^{'} | O_{1} (^{3}S_{1}) | \psi^{'} \rangle
& \approx & 9.7 \times 10^{-2}~~{\rm GeV}^{3},
\end{eqnarray}
in the nonrelativistic limit with $\alpha_{s} (M_{c}) = 0.27$.
\footnote{The radiative
corrections  in $\alpha_s$ has not been included here for consistency.
To be consistent with the velocity counting rules in the
NRQCD in the Coulomb gauge for the heavy quarkonia \cite{bodwin},
one has to include the relativistic corrections as well, since $v \sim
\alpha_{s} (Mv)$  in heavy quarkonium system. If one includes the
$O(\alpha_{s})$ radiative corrections to $J/\psi \rightarrow l^{+} l^{-}$
without relativistic corrections, one gets a larger $\langle 0 |
O_{1}^{J/\psi} (^{3}S_{1}) | 0 \rangle$ compared to the lowest order
result, Eq.~(2.8)
: $ \langle 0 | O_{1}^{J/\psi} (^{3}S_{1}) | 0 \rangle \approx
4.14 \times 10^{-1}~~{\rm GeV}^3$.
Relativistic corrections gives a further
enhancement. See Ref.~[11] for further details.}  From 
these expressions with $M_{b} \approx 5.3$ GeV,
one can estimate the branching ratios for $B$ decays
into $J/\psi + X$ and $\psi^{'} + X$ :
\begin{eqnarray}
B(B \rightarrow J/\psi + X) = 0.23 \%, ~~~~~(0.80 \pm 0.08) \%,
\\
B(B \rightarrow \psi^{'} + X) = 0.08 \%.~~~~~(0.34 \pm 0.04 \pm 0.03) \%.
\end{eqnarray}
The recent data from CLEO \cite{cleo} are shown in the parentheses, where
the cascades from $B \rightarrow \chi_{cJ}(1P) + X$ followed by $\chi_{cJ}
\rightarrow J/\psi + \gamma$ have been subtracted in the data shown.
In view of these results, we may conclude
there are some important pieces missing in the calculations of decay rates
for $B \rightarrow (c\bar{c})_{1} (^{3}S_{1}) + X$ using the color-singlet
model in the nonrelativistic limit.

In view of this, we first estimate the color-octet contributions
to $B \rightarrow
J/\psi +X$, motivated by the suggestion that the color-octet mechanism
might be the solution to the $\psi^{'}$ puzzle at the Tevatron.
Although it is of higher order  in $v^2$ ($\sim O(v^4)$),
it can be important in the case of the inclusive $B$ decays into
$J/\psi + X$, since the Wilson
coefficient of the color-singlet part is suppressed compared to that of the
color-octet part by a factor of $\sim \alpha_{s}$.
( In Eq.~(2.1), $(2C_{+} - C_{-}) \approx 0.4$, and
$(C_{+} + C_{-} ) \approx 2.20$. )
In Ref.~\cite{ko},
a new factorization formula is derived for $B \rightarrow J/\psi + X$ :
\begin{eqnarray}
\Gamma (B \rightarrow J/\psi + X) & = & \left( { \langle 0 | O_{1}^{J/\psi}
(^{3}S_{1}) | 0 \rangle \over 3 M_{c}^2} - 
{ \langle 0 | P_{1}^{J/\psi} (^{3}S_{1}) | 0 \rangle \over 9 M_{c}^4}
\right) ~ (2C_{+} - C_{-}
)^{2} \left( 1 + {8 M_{c}^{2} \over M_{b}^2} \right)~\hat{\Gamma}_{0}
\nonumber        \\
& + & { \langle 0 | O_{8}^{J/\psi}(^{3}S_{1}) | 0 \rangle
\over 2 M_{c}^2}~(C_{+} + C_{-})^{2}~\left( 1 + {8 M_{c}^{2} \over M_{b}^2}
\right)~\hat{\Gamma}_{0}
\nonumber    
\\
& + & {3 \langle 0 | O_{8}^{J/\psi}(^{1}S_{0}) | 0 \rangle
\over 2 M_{c}^2}~(C_{+} + C_{-})^{2}~\hat{\Gamma}_{0} 
\\
& + & { \langle 0 | O_{8}^{J/\psi}(^{3}P_{1}) | 0 \rangle
\over  M_{c}^4}~(C_{+} + C_{-})^{2}~\left( 1 + {8 M_{c}^{2} \over M_{b}^2}
\right)~\hat{\Gamma}_{0},
\nonumber
\end{eqnarray}
with $\Gamma_0$ defined in (2.6).
Using the relations (1.3) and (1.4), we estimate the above branching ratio
to be (for $\alpha_{s}(M_{\psi}^2) = 0.28$ in Ref.~\cite{ko}) 
\begin{equation}
(0.42 \% \times 12.8 )~< B (B \rightarrow J/\psi + X)  < 
~ (0.42 \% \times 13.8) 
\end{equation}
which is larger than the recent CLEO data \cite{cleo} by an order of magnitude 
\footnote{Even if we use the new determination (3.38) by Fleming {\it et al.} 
\cite{fleming2}, we still get a large branching ratio :
\[
(0.42 \% \times 3.45 )~< B (B \rightarrow J/\psi + X)  <
~ (0.42 \% \times 5.45),
\]
although the discrepancy gets milder than the case (2.13).}: 
\begin{equation}
B_{\rm exp} (B \rightarrow J/\psi + X)  = (0.80 \pm 0.08) \%.
\end{equation}
Here, the factor $0.42 \%$ in Eq.~(2.13) 
comes from the color-singlet and the color-octet
${^{3}S_1}$ contributions. Other factor comes from the color-octet
${^{1}S_0}$ and $^{3}P_J$ states, which are very large if one assumes
the relations (1.5) and (1.6).
The situation is the same for $B \rightarrow \psi^{'} + X$.
This is problematic, unless this large color-octet contributions are canceled
by the color-singlet contributions of higher order in $O(\alpha_{s})$ which 
were not included in Ref.~\cite{ko}.   
If there are no such fortuitous cancelations among various color-octet and 
the color-singlet contributions,
this disaster could be attributed to the relation (1.4) being   
too large compared to the naive velocity scaling rule in NRQCD,
as noticed in  Ref.~\cite{cho2}.
It seems to be crucial to include the higher order 
corrections of $O(\alpha_{s}^{4})$ for the color-singlet  $J/\psi$ 
productions at the Tevatron, which is still lacking in the literature.

\section{$J/\psi$ photoproduction}
\label{sec:photo}

The inelastic $J/\psi-$photoproduction has long been studied in the
framework of PQCD and the color-singlet model \cite{berger} \cite{jung}.
The lowest order
subprocess at the parton level for $\gamma + p \rightarrow J/\psi + X$
is the $\gamma-$gluon  fusion at the short distance scale 
\begin{equation}
\gamma + g \rightarrow (c\bar{c})(^{3}S^{(1)}_{1}) + g,
\end{equation}
followed by the long distance process
\begin{equation}
(c\bar{c})(^{3}S^{(1)}_{1}) \rightarrow J/\psi,
\end{equation}
at the  order of $O(\alpha \alpha_{s}^{2} v^{3})$ in the nonrelativistic
limit.
Thus, the production cross section is proportional to the gluon distribution
inside the proton.
This is why  the $J/\psi-$photoproduction has been advocated as a clean
probe for the gluon structure function of a proton in the color-singlet
model.
Without further details, we show the lowest order color-singlet
contribution to $J/\psi$
photoproduction through $\gamma-$gluon fusion in the nonrelativistic limit :
\begin{equation}
\overline{\sum} | {\cal M}(\gamma g \rightarrow J/\psi g) |^{2}
 =  {\cal N}_1~{{ \hat{s}^{2} (\hat{s}-4 M_{c}^{2})^{2} + \hat{t}^{2}
( \hat{t} - 4 M_{c}^{2})^{2} + \hat{u} ( \hat{u} - 4 M_{c}^{2} )^{2} }
\over { (\hat{s}-4 M_{c}^{2})^{2} ( \hat{t} - 4 M_{c}^{2})^{2} ( \hat{u} -
4 M_{c}^{2} )^{2}  }},
\end{equation}
where
\begin{equation}
\begin{array}{ccccl}
    z & = &  \frac{E_\psi}{E_\gamma}|_{\rm lab}&=&
             {p_{N} \cdot P \over p_{N} \cdot k},
\\
\hat{s} &  = & ( k + q_{1} )^{2} &=& x s,
\\
\hat{t} &  = & ( P - k)^{2} &=& (z-1) \hat{s}.
\end{array}
\end{equation}
The overall normalization ${\cal N}_1$ is defined as
\begin{equation}
{\cal N}_1 = {32 \over 9}~( 4 \pi \alpha_{s})^{2} (4 \pi \alpha) e_{c}^{2}
~M_{c}^{3} G_{1}(J/\psi).
\end{equation}
The parameter $G_1 (J/\psi)$, which is defined as
\begin{equation}
G_{1}( J/\psi) =  {\langle J/\psi | {\cal O}_{1} ({^{3}S_{1}}) | J/\psi
\rangle  \over M_c^2}
\end{equation}
in the NRQCD, is proportional to the probability that a color-singlet
$c \bar{c}$ pair in the $^{3}S^{(1)}_{1}$ state to form a
physical  $J/\psi$ state.  It is related with the leptonic decay via
\begin{equation}
\Gamma ( J/\psi \rightarrow l^{+} l^{-} ) = {2 \over 3}~\pi e_{c}^{2}
\alpha^{2}~G_{1}(J/\psi),
\end{equation}
to the lowest order in $\alpha_s$.   From
the measured leptonic decay rate of $J/\psi$, one can extract
\begin{equation}
G_{1}(J/\psi) \approx 106~~{\rm MeV},
\end{equation}
Including the radiative corrections of $O(\alpha_{s})$ with $\alpha_{s}
(M_{c}) =0.27$, it is increased to $\approx 184$ MeV.
Relativistic corrections tend to increase $G_{1} (J/\psi)  $ further to
$\sim 195$ MeV \cite{ko}.
 
The partonic cross section for $\gamma + a \rightarrow J/\psi + b$
is given by
\begin{equation}
{d\hat{\sigma} \over d\hat{t}} = {1 \over 16 \pi \hat{s}^2}~
\overline{\sum}
| {\cal M}(\gamma + a \rightarrow J/\psi + b) |^{2}.
\end{equation}
The double differential cross section is
\begin{equation}
{d^{2}\sigma \over dz dP_{T}^2} (\gamma + p (p_{N}) \rightarrow
J/\psi (P, \epsilon) + X)
= {x g(x,Q^{2}) \over z (1 -z)}~{1 \over 16 \pi \hat{s}^2}~
\overline{\sum}| {\cal M} |^2
(\hat{s}, \hat{t}),
\end{equation}
where
\begin{equation}
x  =  \frac{\hat{s}}{s} =
{1\over z s}~\left[ M_{\psi}^{2} + {P_{T}^{2} \over 1-z } \right].
\end{equation}
One has  the following constraints for $x,z,t$ and $P_{T}^2$ :
\begin{eqnarray}
{M_{\psi}^{2} \over s}    < x < 1,
\\
-(\hat{s} - M_{\psi}^{2}) \leq \hat{t} (=t) \leq 0,
\\
M_{\psi}^{2} \leq {M_{\psi}^{2} \over z} + {P_{T}^{2} \over z(1-z)}
\leq s.
\end{eqnarray}
The $z$ and $P_{T}^2$ distributions can be obtained in the following manner :
\begin{eqnarray}
\frac{d\sigma}{dz}&=&
\int^{(1-z)(zs-M_\psi^2)}_0\frac{d^2\sigma}{dzdP_T^2}dP_T^2,
\\
\frac{d\sigma}{dP_T^2}&=&
\int^{z_{\rm max}}_{z_{\rm min}}\frac{d^2\sigma}{dzdP_T^2}dz,
\\
z_{\rm max}&=&\frac{1}{2s}
\left(
s+M_\psi^2+\sqrt{(s-M_\psi^2)^2-4sP_T^2}
\right),
\\
z_{\rm min}&=&\frac{1}{2s}
\left(
s+M_\psi^2-\sqrt{(s-M_\psi^2)^2-4sP_T^2}
\right).
\end{eqnarray}
 
There are two kinds of corrections to the lowest order result in the
color-singlet model (3.3) : the relativistic corrections of $O(v^{2})$
and the PQCD radiative corrections of $O(\alpha_{s})$ relative to the
lowest order result shown in (3.3).
We briefly summarize both types of corrections 
here, since they have to be included in principle
for consistency, when one
includes the color-octet mechanism in many cases.
 
The relativistic corrections to the $\gamma-$gluon fusion were studied
by Jung {\it et al.} \cite{jung}. They found that relativistic corrections
are important for high $z > 0.9$ at EMC energy ($\sqrt{s_{\gamma p}} \simeq
14.7$ GeV). Since it mainly affects the high $z$ region only, we neglect
the relativistic corrections, keeping in mind that it enhances the cross
section at large $z > 0.9$.
 
The radiative corrections to the $J/\psi$ photoproduction is rather
important in practice.
This calculation has been done recently in Ref.~\cite{kramer}, and
the scale dependence of the lowest order result ($Q^{2}$ in the structure
function in Eq. (3.10) ) becomes
considerably reduced. For EMC energy region, the $K$ factor is rather large,
$K \sim 2$. For HERA, it depends on the cuts in
$z$ and $P_{T}^2$.
We include the radiative corrections in terms of a $K$ factor suitable
to the energy range we consider.
Another consequence of the radiative corrections to
the color-singlet $J/\psi$ photoproduction is that the PQCD
becomes out of control for $z > 0.9$ at EMC energy. For HERA, one gets
reasonable results in PQCD when one imposes the following cuts in $z$ and
$P_{T}^{2}$ : $z < 0.8$ and $P_{T}^{2} > 1~{\rm GeV}^2$ .
Thus, it does not make much sense to talk about
the $z$ or $p_T$ distributions for such $z$ region in PQCD.
One has to introduce cuts in $z$ as well as in $p_T$. Following
the Ref.~\cite{kramer}, we adopt the following sets of cuts :
\begin{eqnarray}
z < 0.9, & ~~~~~& {\rm for ~EMC},
\\
0.2 < z < 0.8 & ~~~~~& {\rm for ~HERA}.
\end{eqnarray}
At HERA energies, the lower cut in $z (z > 0.2)$ is employed in
order to reduce  backgrounds from the
resolved photon process and the $b$ decays into $J/\psi$.
For these cuts, the $K$ factor is approximately $K \simeq 1.8$ both
at HERA and the fixed target experiments.
We include these radiative corrections to the
subprocess (3.1)  by setting $K \simeq 1.8$.

Let us consider color-octet contributions to the $2 \rightarrow 1$
subprocesses via
\begin{equation}
\gamma(k) + g^*_a(g) \rightarrow (c\bar{c})[^{2S+1}L^{(8b)}_J](P),
\end{equation}
followed by $(c\bar{c})_{8}$ fragmenting into $J/\psi$ with emission of
soft gluons.
This subprocess occurs at $O(\alpha \alpha_{s} v^{7})$.
Here, $a, b$ are color indices for the initial gluon and the final
color-octet $c\bar{c}$ state, and we are interested in $S=L=J=0$ and
$S=L=1, J=0,1,2$.
There are 2 diagrams representing the vertex.
Here we consider the process where only the gluon is  off-shell.
Following the conventions adopted in the previous section,
we first write the matrix ${\cal O}$ related to this effective vertex.
\begin{eqnarray}
{\cal O}(P,q)
&=&
\frac{ee_cg_s\delta^{ab}}{\sqrt{2}}
\left[
\not{\epsilon}^\gamma
\frac {\frac{\not{P}}{2}+\not{q}-\not{k}+M_c}
{(\frac{P}{2}+q-k)^2-M_c^2}
\not{\epsilon}^g
+
\not{\epsilon}^g
\frac {\frac{\not{P}}{2}+\not{q}-\not{g}+M_c}
{(\frac{P}{2}+q-g)^2-M_c^2}
\not{\epsilon}^\gamma
\right].
\end{eqnarray}
With this matrix ${\cal O}$
we can derive the effective vertices for the
$\gamma g (c\bar{c})^{2S+1}L_J^{(8)}$
as
\begin{eqnarray}
{\cal M}^\prime(^1S_0^{(8)})&=& 4i\frac{ee_cg_s}{g^2-4M_c^2}
\delta^{ab} \epsilon^{\mu\nu\kappa\lambda}
\epsilon^\gamma_\mu \epsilon^g_\nu P_\kappa k_\lambda,
\\
{\cal M}^\prime(^3S_1^{(8)})&=& 0,
\\
{\cal M}^\prime(^3P_0^{(8)})&=&
\frac{2e e_c g_s \delta^{ab}}{\sqrt{3}M_c}
\left(\frac{g^2-12M_c^2}{g^2-4M_c^2}\right)
\left(g^{\mu\nu}+2\frac{P^\mu k^\nu}{g^2-4M_c^2}\right)
\epsilon^\gamma_\mu
\epsilon^g_\nu,
\\
{\cal M}^\prime(^3P_1^{(8)})&=&
\frac{\sqrt{2}e e_c g_s\delta^{ab}}{M_c^2(g^2-4M_c^2)}
\nonumber\\
&\times &
\left(
      g^2\epsilon^{\mu\nu\alpha\tau}
     +{2k_\kappa}
       \frac{ g^2
               (  P^\mu\epsilon^{\nu\alpha\kappa\tau}
                 -P^\nu\epsilon^{\mu\alpha\kappa\tau}
                )
        +4g^\nu M_c^2\epsilon^{\mu\alpha\kappa\tau}
       } {g^2-4M_c^2}
\right)
\epsilon_{\alpha}(J_z)
\epsilon^\gamma_\mu
\epsilon^g_\nu
P_\tau,
\\
{\cal M}^\prime(^3P_2^{(8)})&=&
\frac{16e e_c g_s\delta^{ab}}{(g^2-4M_c^2)}M_c
\left(
g^{\mu\alpha} g^{\nu\beta}
+2k^\alpha
\frac{
k^\beta g^{\mu\nu} + P^\mu g^{\nu\beta} - k^\nu g^{\mu\beta}
}
{g^2-4M_c^2}
\right)
\epsilon_{\alpha\beta}(J_z)
\epsilon^\gamma_\mu
\epsilon^g_\nu.
\end{eqnarray}
Since $J/\psi$ can be produced
via the $2\rightarrow 1$ subprocesses with these effective vertices,
we can obtain the $2\rightarrow 1$ color octet contribution
by using the following average squared amplitudes as \footnote{Our results
agree with those obtained in Refs.~\cite{kramer2} \cite{fleming2}.
Note, however, that our convention of the invariant matrix is different from
theirs.}
\begin{eqnarray}
\overline{\sum}|{\cal M}^\prime(^{1}S_0^{(8)})|^2&=&2(ee_cg_s)^2,
\\
\overline{\sum}|{\cal M}^\prime(^{3}S_1^{(8)})|^2&=&0,
\\
\overline{\sum}|{\cal M}^\prime(^{3}P_0^{(8)})|^2&=&\frac{6}{M_c^2}
(ee_cg_s)^2,
\\
\overline{\sum}|{\cal M}^\prime(^{3}P_1^{(8)})|^2&=&0,
\\
\overline{\sum}|{\cal M}^\prime(^{3}P_2^{(8)})|^2&=&
\frac{8}{M_c^2}(ee_cg_s)^2.
\end{eqnarray}
The $J/\psi$ photoproduction cross section
via $2\rightarrow 1$ process is given by 
\begin{eqnarray}
&&\sigma\left(\gamma+p\rightarrow (c\bar{c})^{(8)}\rightarrow \psi\right)
\nonumber\\
&&\hskip 2cm
=\frac{7\pi(ee_cg_s)^2}{64M_c^5}\left[xf_{g/p}(x)\right]_{x=4M_c^2/s}
\left(
      \frac{\langle 0|{\cal O}^{\psi}(^3P_0^{(8)})|0\rangle}{M_c^2}
     +\frac{\langle 0|{\cal O}^{\psi}(^1S_0^{(8)})|0\rangle}{7}
\right).
\end{eqnarray}
Since $\hat{\sigma} \propto \delta ( 1 -z)$, this $2\rightarrow 1$
color-octet subprocesses contribute to the elastic peak in the
$J/\psi-$photoproduction.  It is timely to recall that the color-singlet
model with relativistic corrections still underestimates the cross section
for $z \ge 0.9$ by an appreciable amount \cite{jung}.
As $z \rightarrow 1$, the final state gluon in the $\gamma-$gluon fusion
becomes softer and softer, although this does not cause any infrared
divergence in the transition matrix element.
Therefore, it would be more meaningful to factorize the effect of this
final soft gluon
into the color-octet matrix elements, $\langle O_{8}^{\psi}
(^{1}S_{0}) \rangle$ and  $\langle O_{8}^{\psi} (^{3}P_{J} ) \rangle$.
The color-octet $^{1}S_{0}$ and
$^{3}P_{J}$ states might reduce the gap between the color-singlet prediction
and the experimental value of $d\sigma / dz$ for $0.9 \le z \le 1$.

The color-octet $2 \rightarrow 1$ subprocess (3.21) considered 
before not only contributes to the elastic peak of the $J/\psi$
photoproduction, but it also contributes
to the resolved photon processes at $O(\alpha \alpha_{s}^{2} v^{7})$,
where the initial partons
can be either gluon or light quarks ($q = u,d,s$).
These processes are suppressed by $v^4$ but enhanced by $1/\alpha_s$, relative
to the resolved photon process in the color-singlet model. Also, 
they can enhance the high-$p_T
J/\psi$'s, which might be relevant to the $J/\psi$ photoproduction
at HERA.
 This can be a background to the
determination of gluon distribution function of a proton, if the cross
section is appreciable.  The resolved photon process in the color-singlet
model is dominant over the $\gamma-$gluon fusion in the lower $z$ region,
$z < 0.2$, and it can be discarded by a suitable cut on $z$.
Since the color-octet contribution to the resolved photon process has not
been studied in the literature, we address this issue here.
When one considers 
\begin{eqnarray*}
\gamma + g & \rightarrow & (c\bar{c})_{8} ({^1S_0}~{\rm or} ~ {^3P_J}),
\\
\gamma + q & \rightarrow & (c\bar{c})_{8} ({^1S_0}~{\rm or} ~ {^3P_J}),
\end{eqnarray*}
one has to include 
\[
\gamma + g \rightarrow (c\bar{c})_{8} ({^3S_1})
\]
simultaneously,
since both are the same order of $O(\alpha \alpha_{s}^{2} v^{7})$.
This diagram is the same as the color-singlet case except for the color
factor of the $(c \bar{c})$ state.
 
It is straightforward, although lengthy,
to calculate the amplitudes for the above three processes.
Using REDUCE in order to the spinor algebra in a symbolic manner,
we can get the averaged ~${\cal M}$~ squared for various ~$2\to 2$~processes.
 
Another  color-octet $(c\bar{c})(^{3}S^{(8)}_{1})$ contribution to the
$J/\psi-$photoproduction comes from the Compton scattering type subprocesses
 :
\begin{equation}
\gamma (k,\epsilon) + q(p_{1}) \rightarrow (c\bar{c})(^{3}S^{(8a)}_{1})
(P,\epsilon^{*}) + q(p_{2}),
\end{equation}
where $P$ and $\epsilon^{*}$ are the four momentum and the polarization vector
of the  $^{3}S_{1}$ color-octet state, and $a$ is its color index.
This subprocess, if important, can be a background to the determination of
the gluon distribution function in a proton, since it is initiated by
light quarks. From
the naive power counting, however, we infer this subprocess occurs at
$O(\alpha \alpha_{s}^{2})$ in the coupling constant expansion, and also
suppressed by $v^4$ compared to the color-singlet contribution (3.1)
due to its color-octet nature.  Thus, this subprocess is expected to be
negligible.
 
One can actually quantify this argument by explicitly evaluating the
Feynman diagrams for (3.34).
The effective vertex for $q \bar{q} \rightarrow
(c\bar{c})(^{3}S^{(8a)}_{1})$
is given by  \cite{cho}
\begin{equation}
{\cal M}^\prime
(q(p_{1}) \bar{q} (p_{2}) \rightarrow (c\bar{c})(^{3}S^{(8a)}_{1})) =
{4 \pi \alpha_{s} \over 2 M_{c}}~\bar{v}(p_{2}) \gamma^{\mu} T^{a} u(p_{1})~
\epsilon_{\mu}^{*}(p_{1}+p_{2}, S_{z}),
\end{equation}
where $\epsilon_{\mu}^{*}$ is the polarization of the produced spin-1 color
octet object.
Using this effective vertex, one can calculate the amplitude for 
$\gamma q \rightarrow (c\bar{c})_8 ({^3S_1}^{(8a)}) q$  
\begin{eqnarray}
{\cal M}^\prime
(\gamma q \rightarrow (c\bar{c})(^{3}S^{(8a)}_{1}) q) & = &
-{g_{s}^{2} e e_{q} \over 2 M_c}~
\bar{u}(p_{2}) ~
\left[ \not{\epsilon}^{*}(P,S_{z}) T_{a} ~{(k+p_{1}+M_{c}) \over
(k+p_{1})^{2} -M_{c}^2}~\not{\epsilon}_\gamma
\right.   \nonumber\\
&&\hskip 3cm + \left.   \not{\epsilon}_{\gamma} {(p_{1}-P + M_{c})
\over  (p_{1}-P)^{2} - M_{c}^2}
\not{\epsilon}^{*}(P,S_{z}) T_{a} \right]~u(p_{1}).
\end{eqnarray}
where $ee_q$ is the electric charge of the light quark
inside proton($q=u, d, s$).
The average amplitude squared for the color-octet $^{3}S_1$ state is given by
\begin{equation}
\overline{\sum}| {\cal M}^\prime(\gamma q \rightarrow
(c\bar{c})(^{3}S^{(8)}_{1}) q) |^{2}
=  -\frac{2}{3M_c^2}(g_s^2 e e_q)^2
(\frac{\hat{s}}{\hat{u}}+ \frac{\hat{u}}{\hat{s}}+8\frac{M_c^2
\hat{t}}{\hat{s} \hat{u}}).
\end{equation}
This completes our discussions on the color-octet $2 \rightarrow 2$
subprocess for $J/\psi$ photoproductions.
 
Now, we  are ready to show the numerical results using the analytic
expressions obtained in the previous section.  Let us first summarize the
input parameters and the structure functions we will use in the
following.  The results are quite sensitive to the numerical values of
$\alpha_s$ and $m_{c}$ and the factorization scale $Q$.
We shall use $\alpha_{s} (M_{c}^{2}) = 0.3$, $m_{c} = 1.48$ GeV and
$Q^{2} = (2 m_{c})^2$. For the structure functions, we use
the most recent ones, MRSA \cite{mrsa} and CTEQ3M \cite{cteq3},
which incorporate the new data
from HERA \cite{exp:hera}, on the lepton asymmetry in $W-$boson production
\cite{exp:lepton} and on the difference in Drell-Yan cross sections
from proton and neutron targets \cite{exp:DY}.
For the $2 \rightarrow 1$, we show results using both structure functions.
For the $2 \rightarrow 2$ case, we show the results with the CTEQ3M structure
functions only, since the MRSA structure functions yield almost the
same results within $\sim 10 \%$ or so.
 
Let us first consider the $J/\psi$ photoproduction via the color-octet
$2 \rightarrow 1$ subprocess. 
Since the subprocess cross section (3.33) vanishes except at $z=1$,
one can infer that it contributes to the $J/\psi$ photoproductions
in the forward direction ($z \sim 1, P_{T}^{2} \simeq 0$).
In Figs.~1 (a) and (b), we show the $J/\psi$ photoproduction cross section
in the forward direction ($\sigma_{forward}$)
as well as the data from the fixed target
experiments and the preliminary data from H1 at HERA, respectively.
In each case, the upper and the lower curves define the region allowed
by the relation (1.4) for two color-octet matrix elements,
$\langle 0 | O_{8}^{\psi} ({^{3}S_{1}}) | 0 \rangle$ and $\langle 0
| O_{8}^{\psi} ({^{3}P_{0}}) | 0 \rangle$.
In case of fixed target experiments, $\sigma_{forward}$   is
usually characterized by $z>0.9$, with the remainder being associated
with the inelastic $J/\psi$ photoproduction.  According to this criterion,
the experimental value of $\sigma_{\rm exp} (\gamma + p \rightarrow J/\psi
+ X)$ contains contributions from inelastic production of $J/\psi$'s.
Thus, the data should lie above the predictions from the color-octet
$2 \rightarrow 1$ subprocess, (3.21).  Fig.~1 (a) shows that the situation
is opposite to this expectation.  Color-octet contributions are larger
than the data, which indicates that the numerical values of the color-octet
matrix elements are probably too large.
At HERA, one has the elastic $J/\psi$ photoproduction data, which can be
identified with  the color-octet $2 \rightarrow 1$ subprocess.
By saturating the relation
(1.4) by either color-octet matrix element, we get the
$J/\psi$ photoproduction cross section in the forward direction
(Fig.~1 (b)).  We observe again that the color-octet contribution with (1.4)
overestimates the cross section by a large amount.
This disagreement can arise from two sources : (i) the radiative
corrections to $p \bar{p} \rightarrow J/\psi + X$, which were ignored in
Ref.~\cite{cho2} is important, and/or (ii) the heavy quark spin symmetry for
$\langle 0 | O_{8}^{\psi} ({^{3}P_{J}}) | 0 \rangle \approx
(2J+1)~\langle 0 | O_{8}^{\psi} ({^{3}P_{0}}) | 0 \rangle$ may not be
a good approximation. Although the heavy quark spin symmetry relation
is used quite often in heavy quarkonium physics, it may be violated
by a considerable amount \cite{ko}.
 
Recently, Amundson  {\it et al.} performed the $\chi^2$ fit to the
available fixed target experiments and the HERA data independently, and
found that [13]
\begin{equation}
\langle 0 | O_{8}^{\psi} (^{1}S_{0}) | 0 \rangle + {7 \over M_{c}^2}~
\langle 0 | O_{8}^{\psi} (^{3}P_{0}) | 0 \rangle  =
(0.020 \pm 0.001)~{\rm GeV}^3,
\end{equation}
using the MRSA$^{(')}$, and CTEQ3M structure functions with $\alpha_{s}
(2 M_{c}) = 0.26$ and $M_{c} = 1.5$ GeV.
This determination is not compatible with the relation (1.4), since the
resulting $\langle O_{8}(^{3}P_{0}) \rangle$ is negative.
This is another way to say that the determination of the color-octet
matrix elements from the $J/\psi$ productions at the Tevatron may not be
that reliable. In fact, this is not very surprising, since the radiative
corrections to the lowest-order color-singlet contributions to
the $J/\psi$ hadroproductions are not included yet.
 
Next, we consider the $J/\psi$ photoproduction through $2 \rightarrow 2$
parton-level subprocesses.
As discussed before, the PQCD corrections
to the lowest order $\gamma + g \rightarrow J/\psi + g$ is not under
proper control for $z> 0.9$. Therefore, we impose a cut $z < 0.9$ at
EMC energy, $\sqrt{s_{\gamma p}} = 14.7$ GeV,  and
at HERA with $\sqrt{s_{\gamma p}} = 100$ GeV,
we impose cuts on $z$ and $P_{T}^2$ \cite{kramer} :
\[
0.2 < z < 0.8, ~~~~~~~~~~~~~~~~~~~P_{T}^{2} > 1~{\rm GeV}^2.
\]
In both cases, we set $K \simeq 1.8$.
 
In Figs.~2 (a) and (b), we show the $d\sigma / dz$  distributions of
$J/\psi$ at EMC (NMC) and HERA along with the corresponding data.
In both cases, the color-octet ${^3S_1}$ contribution
(Compton scattering type) is negligible in most regions of $z$, and thus
can be safely neglected.
The thick dashed and the thin dashed  curves correspond to the cases where
the relation (1.4) is saturated by $\langle 0 | O_{8}^{\psi}(^{3}P_{J}) | 0
\rangle$ and $\langle 0 | O_{8}^{\psi} (^{1}S_{0}) | 0 \rangle$,
respectively.  The thick and the thin solid curves
represent the sum of the color-singlet and the color-octet contributions,
in case that the relation (1.4) is saturated by $\langle 0 |
{\cal O}_{8}^{\psi}
({^1S_0}) | 0 \rangle$  and  $\langle 0 | {\cal O}_{8}^{\psi}({^3P_0}) | 0
\rangle$, respectively. In either case, we observe that
the color-octet ${^1S_0}$ and ${^3P_J}$ contributions begin to dominate the
color-singlet contributions for $z > 0.6$, and become too large for
high $z$ region considering we have not added the enhancements at high $z$
due to the relativistic corrections.
Thus, this behavior of rapid growing at high $z$ does not agree with the
data points at EMC and HERA, if we adopt the determination (1.4) by Cho and
Leibovich  \cite{cho2}.
 
In Fig.~3, we show the inelastic $J/\psi$ photoproduction cross section as
a function of $\sqrt{s_{\gamma p}}$ with the cut, $ z < 0.8$ and
$P_{T}^{2} > 1~{\rm GeV}^2$.  Again, the color-octet ${^3S_1}$ contribution is
too small, and thus not shown in the figure.
Here again, the color-octet ${^1S_0}$ and ${^3P_J}$ contributions via
$2 \rightarrow 2$ subprocesses  dominate the color-singlet
contribution, if the relation (1.4) is imposed.  Although the total seems
to be in reasonable agreement with the preliminary H1 data, direct comparison
may be meaningful only if the cascade $J/\psi$'s from $b$ decays have been
subtracted out.  There are also considerable amount of uncertainties coming
from $M_c$ and $\alpha_s$.  Therefore, it is sufficient to say that the
color-octet ${^1S_0}$ and ${^3P_J}$ states dominate the singlet contribution
to the $J/\psi$ photoproduction, if the relation (1.4) is imposed.

\section{Conclusion}
\label{sec:con}

In summary, we considered the color-octet contributions to (A) the inclusive 
$J/\psi$ productions in $B$ decays, and (B) the $J/\psi$ photoproductions
($\gamma  + p \rightarrow J/\psi + X$) through (i) $\gamma g \rightarrow
(c\bar{c})_{8} ({^1S_0}~{\rm and}~{^3P_J})$  and the subsequent evolution of 
$(c\bar{c})_8$  into a physical $J/\psi$  with $z \approx 1$ and $P_{T}^2 
\approx 0$, (ii) the subprocesses $\gamma + g ({\rm or}~q) \rightarrow 
(c\bar{c})_{8} ({^1S_0}~{\rm or}~{^3P_J}) + g ({\rm or}~ q)$.  
These are compared with (i) the measured $J/\psi$ photoproduction cross 
section in the forward direction, and (ii) the $z$ distributions of $J/\psi$ 
at  EMC and HERA, and the preliminary result on the inelastic $J/\psi$ 
photoproduction total cross section at HERA. One finds that the relation (1.4) 
color-octet   lead to too large contributions of the color-octet ${^1S_0}$ and 
${^3P_J}$ states to the above observables.  Especially, the first two 
observables contradict the observation.   This is also against the naive
expectation that the color-octet contribution may not be prominent as in the
case of the $J/\psi$ hadroproductions, since they are suppressed by $v^4$
(although enhanced by one power of $\alpha_s$) relative to the color-singlet
contribution.  
It is also pointed out that the same is true of the process $B \rightarrow
J/\psi +X$, in which the relation (1.4) predicts its branching ratio to be
too large by an order of magnitude compared with the data.

Therefore, one may conclude  that the color-octet matrix
elements involving $c\bar{c}_{8}({^{1}S_{0}}, {^{3}P_{J}})$ might be 
overestimated by an order of magnitude.  Since the relation (1.4) has been 
extracted by fitting the $J/\psi$ production at the Tevatron to the lowest
order color-singlet and the color-octet contributions,
it may be changed when one considers the radiative corrections to the 
lowest order color-singlet contributions.

\acknowledgements

The author is grateful to Prof. H.S. Song and Mr. Jungil Lee for enjoyable
collaborations on the subjects discussed in this article.  
This work was supported in part by KOSEF through CTP at Seoul National 
University, and by the Basic Science Research Program, Ministry of 
Education, 1996, Project No. BSRI--96--2418.


%
%

{\bf Figure Captions}
\\
\\
Fig.1(a)
\hskip .3cm
{
The cross sections for $\gamma + p \rightarrow J/\psi + X$
in the forward direction at
the fixed target experiments
as a function of $E_{\gamma}$.
The solid and the dashed curves were obtained using the CTEQ3M and the MRSA
structure functions.
Here, TOT$_s$ is the $^1S^{(8)}_0$ saturated curve and
TOT$_p$ is the $^3P^{(8)}_J$ saturated one.
}
\\
\\
Fig.1(b)
\hskip .3cm
{
The cross sections for $\gamma + p \rightarrow J/\psi + X$
in the forward direction at
HERA
as a function of the square root of $s_{\gamma p}$.
The solid and the dashed curves were obtained using the CTEQ3M and the MRSA
structure functions.
Here, TOT$_s$ is the $^1S^{(8)}_0$ saturated curve and
TOT$_p$ is the $^3P^{(8)}_J$ saturated one.
}
\\
\\
Fig.2(a)
\hskip .3cm
{
The differential cross sections $d\sigma / dz$ for
$\gamma + p \rightarrow J/\psi + X$  at EMC
as a function of $z\equiv E_{J/\psi}/ E_\gamma$.
The singlet contributions are in the thick dotted curve,
the color-octet $^1S_0$ contributions in the thick dashed curve
(with $\langle O_{8}^{\psi} (^1S_0) \rangle = 6.6 \times 10^{-2}
~{\rm GeV}^{3}$),
 and the color-octet $^3P_J$ contributions in the thin dashed curve
(with $\langle O_{8}^{\psi} (^3P_J) \rangle
/M_c^2= 2.2 \times 10^{-2}
~{\rm GeV}^{3}$).
 The total is shown in the solid curve.
The relation (1.4) allows the region between two solid curves.
Here, TOT$_s$ is the $^1S^{(8)}_0$ saturated curve and
TOT$_p$ is the $^3P^{(8)}_J$ saturated one.
}
\\
\\
Fig.2(b)
\hskip .3cm
{
The differential cross sections $d\sigma / dz$ for
$\gamma + p \rightarrow J/\psi + X$  at HERA
as a function of $z\equiv E_{J/\psi}/ E_\gamma$.
The singlet contributions are in the thick dotted curve,
the color-octet $^1S_0$ contributions in the thick dashed curve
(with $\langle O_{8}^{\psi} (^1S_0) \rangle = 6.6 \times 10^{-2}
~{\rm GeV}^{3}$),
 and the color-octet $^3P_J$ contributions in the thin dashed curve
(with $\langle O_{8}^{\psi} (^3P_J) \rangle
/M_c^2= 2.2 \times 10^{-2}
~{\rm GeV}^{3}$).
 The total is shown in the solid curve.
The relation (1.4) allows the region between two solid curves.
Here, TOT$_s$ is the $^1S^{(8)}_0$ saturated curve and
TOT$_p$ is the $^3P^{(8)}_J$ saturated one.
}
\\
\\
Fig.3
\hskip .3cm
{ Total inelastic $J/\psi$ photoproduction cross section for $z < 0.8$
as a function of the square root of $s_{\gamma p}$.
The singlet contributions in the thick dotted curve,
the color-octet $^1S_0$ contributions in the thick dashed curve
(with $\langle O_{8}^{\psi} (^1S_0) \rangle = 6.6 \times 10^{-2}
~{\rm GeV}^{3}$),
 and the color-octet $^3P_J$ contributions in the thin dashed curve
(with $\langle O_{8}^{\psi} (^3P_J) \rangle
/M_c^2= 2.2 \times 10^{-2}
~{\rm GeV}^{3}$).
 The total is shown in the solid curve.
The relation (1.4) allows the region between two solid curves.
Here, TOT$_s$ is the $^1S^{(8)}_0$ saturated curve and
TOT$_p$ is the $^3P^{(8)}_J$ saturated one.
}

%
%

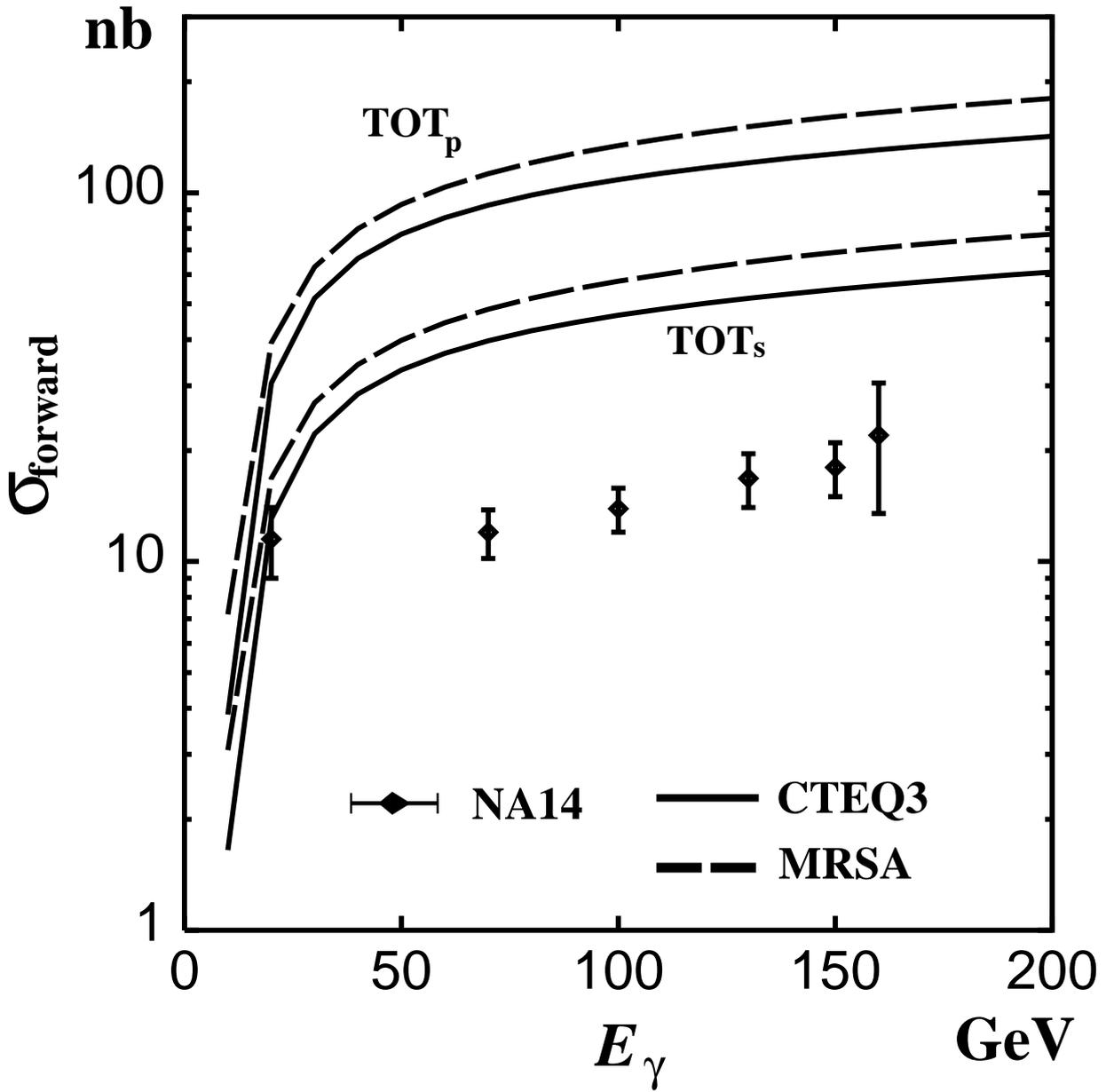
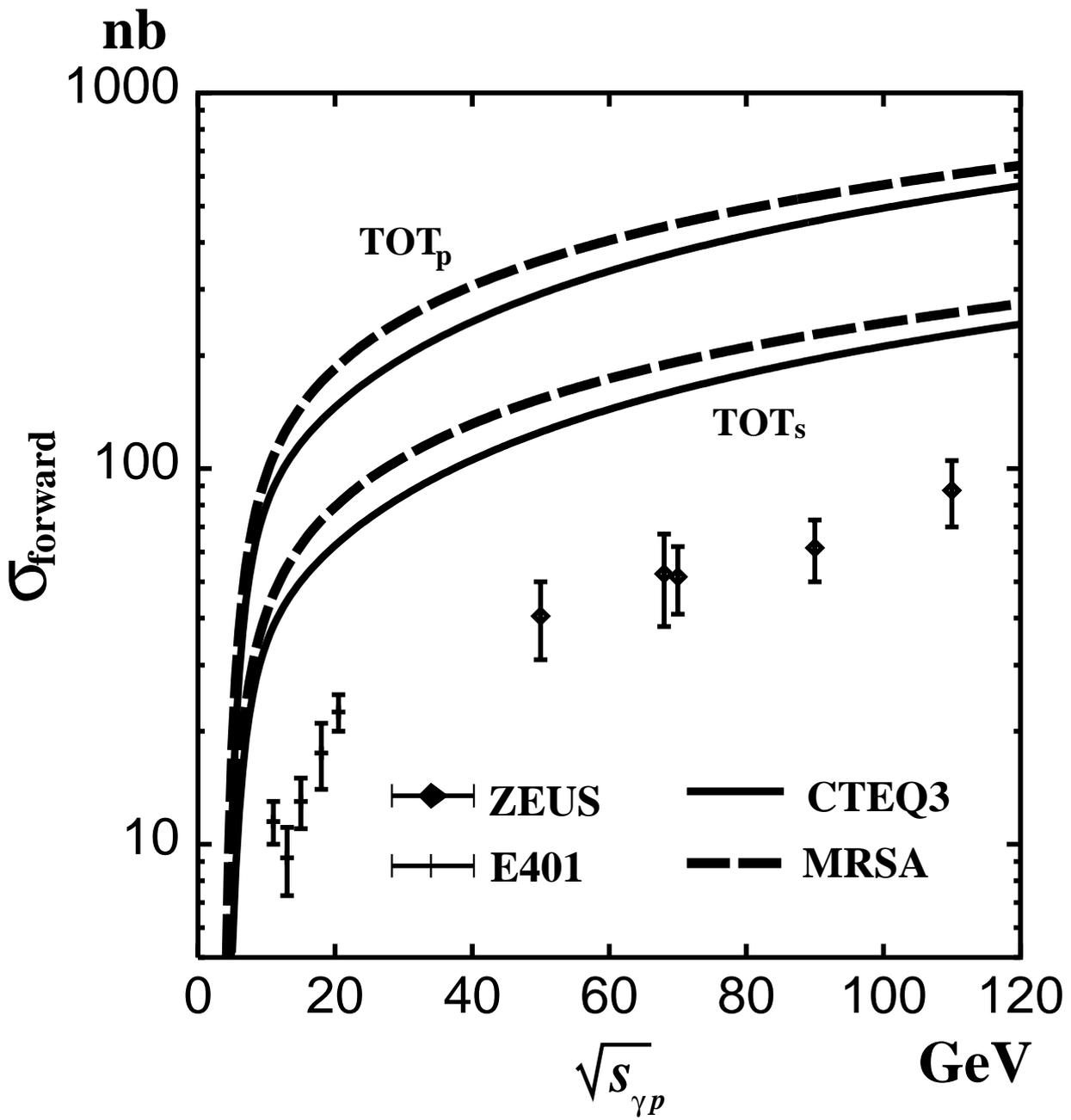
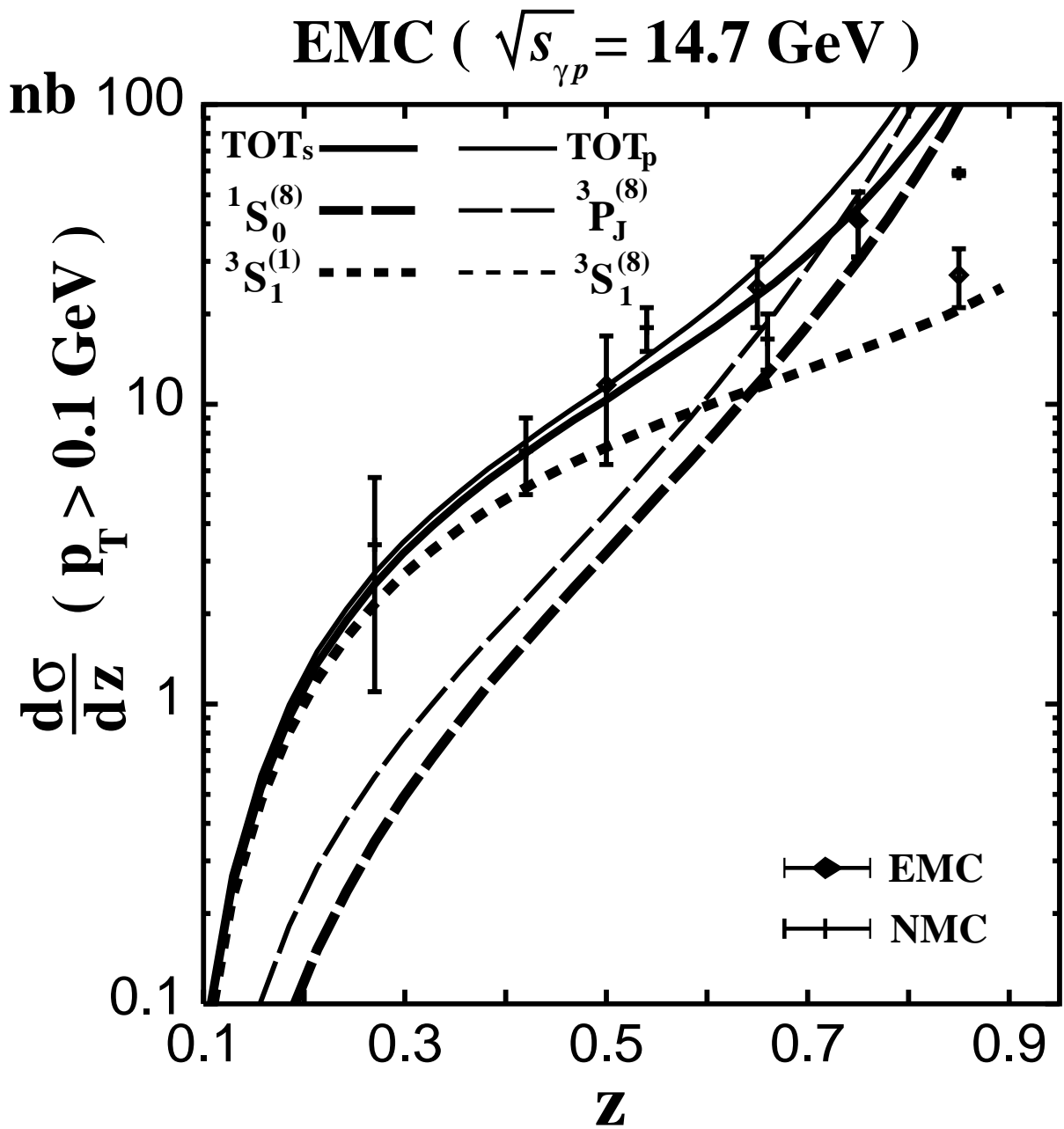
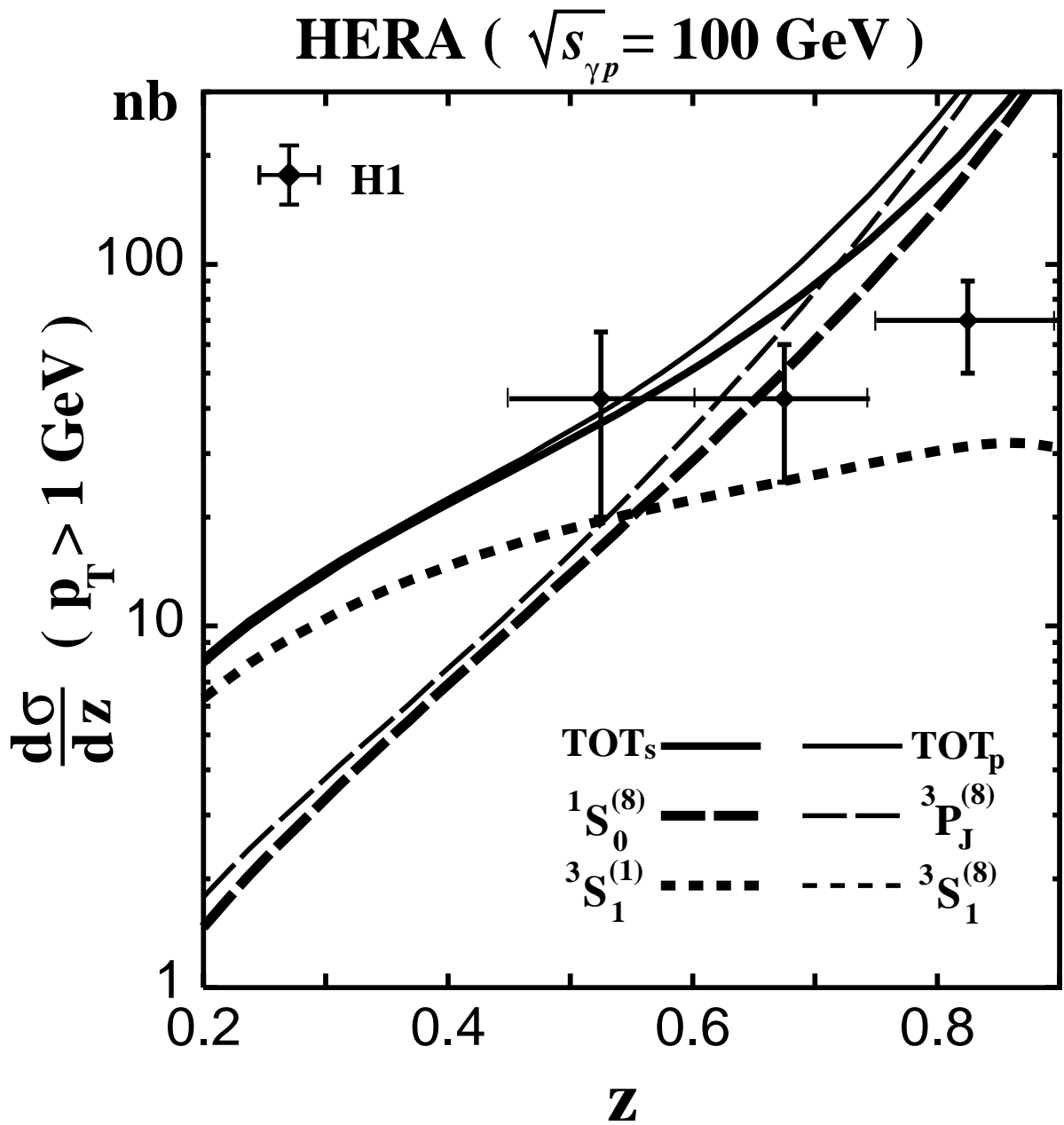
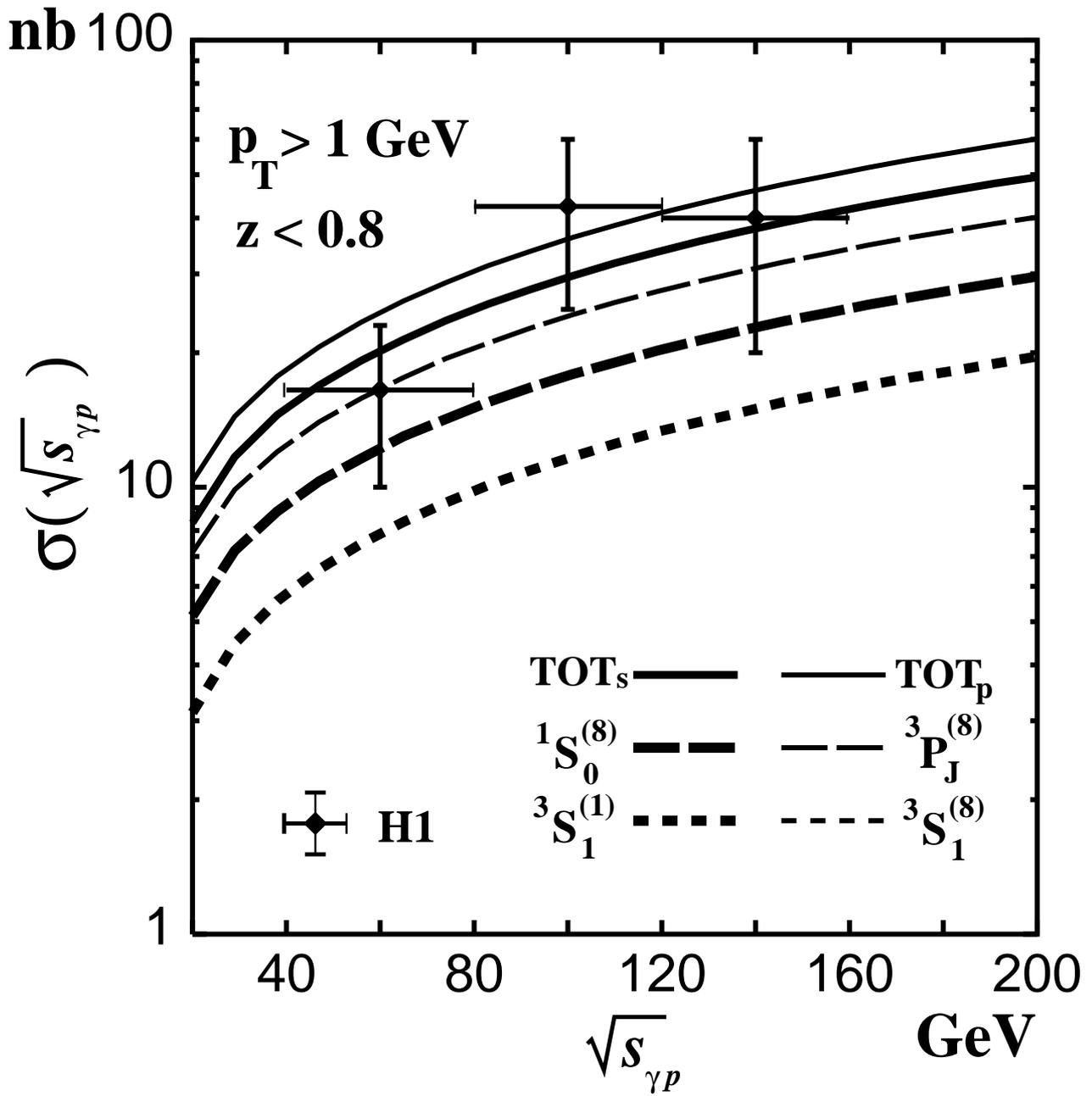
\end{document}